\documentclass[11pt]{article}
\usepackage{blois,epsfig}

\def\PRL{{\em Phys. Rev. Lett.}}
\def\PRD{{\em Phys. Rev.} D}

\def\be{\begin{equation}}
\def\ee{\end{equation}}
\def\bea{\begin{eqnarray}}
\def\eea{\end{eqnarray}}

\begin{document}
\vspace*{4cm}
\title{ THE DENSITY PROFILE OF MASSIVE GALAXY CLUSTERS FROM WEAK LENSING}

\author{ H. DAHLE }

\address{Institute of Theoretical Astrophysics, University of Oslo,\\
P.O. Box 1029, Blindern, N-0315 Oslo, Norway}

\maketitle\abstracts{
We use measurements of weak gravitational shear around a sample of massive galaxy 
clusters at $z = 0.3$ to constrain their average radial density profile.
Our results are consistent with the density profiles of CDM halos in numerical 
simulations and inconsistent with simple models of self-interacting dark matter. 
Unlike some other recent studies, we are not probing the 
scales where the baryonic mass component becomes dynamically important, and so our results
should be directly comparable to CDM N-body simulations.}

\section{Introduction}
While the concordance flat $\Lambda$CDM model, in which the matter density is 
dominated by cold dark matter (CDM), provides a good fit to observed large 
scale-properties of the universe, there remain some possible small-scale 
problems for this model. 

Numerical simulations of structure formation in a CDM model 
predict that the dark matter (DM) halos of $L_{\star}$ galaxies such as the 
Milky Way should contain a number of subhalos that exceed the 
observed number of satellite dwarf galaxies by 1-2 orders of magnitude
(e.g. Klypin et al.\ 1999; Moore et al.\ 1999a). Strongly suppressed star 
formation in the subhalos could be a possible solution to this problem. 
Observations of anomalous flux ratios of strongly 
gravitationally lensed multiple quasar images (Kochanek \& Dalal 2003) and 
observations of the dynamics of optically dark high-velocity gas clouds in the local 
group (Robishaw, Simon \& Blitz 2002) appear to be qualitatively
consistent with this proposed solution. 

In addition, the simulations predict that DM halos have cuspy inner 
density profiles $\rho (r) \propto r^{-\alpha}$, with $\alpha$ somewhere in the 
range between $1.0$ (Navarro, Frenk \& White 1997; hereafter NFW) and $1.5$ 
(Moore et al.\ 1999b). This appears to contradict the observed dynamics of 
DM-dominated low surface brightness galaxies which favour softer cores with 
$\alpha = 0.2 \pm 0.2$ (de Blok, Bosma, \& McGaugh 2003). 
On the scales of galaxy clusters, some studies indicate shallower density 
profiles than those predicted from CDM simulations (Sand et al.\ 2003), while 
others give $\alpha$ values that are consistent with CDM predictions (Bautz \& 
Arabadjis 2003). 

Attempts have been made to solve these small-scale problems of CDM by 
proposing DM models that modify its behavior on small scales. 
Some examples of these are models in which the DM is 
self-interacting (Spergel \& Steinhardt 2000), self-annihilating (Kaplinghat, 
Knox \& Turner 2000), fluid (Peebles 2000; Arbey, Lesgourgues \& Salati 2003), 
warm (e.g., Sommer-Larsen \& Dolgov 2001), repulsive (Goodman 2000), 
fuzzy (Hu, Barkana \& Gruzinov 2000), decaying (Cen 2001), is both 
self-interacting and warm (Hannestad \& Scherrer 2000), acts as 
mirror matter (Mohapatra, Nussinov \& Teplitz 2002) or has its gravitational 
interaction with baryonic matter suppressed on small scales (Piazza \& Marioni 2003). 
Of these, the self-interacting DM model of Spergel \& Steinhardt is the one
which has been explored in most detail. Here, we put limits on this model by 
using weak gravitational lensing to measure the average density profile of 
an ensemble of massive galaxy clusters. Details of this work are given by Dahle, Hannestad \& 
Sommer-Larsen (2003).    

\section{Constraints on the DM halo profile}\label{subsec:constr}

Our data set is a subset of the weak gravitational lensing measurements of 
38 X-ray luminous 
clusters presented by Dahle et al.\ (2002). This subset consists of 6 clusters 
at $z=0.3$ for which weak gravitational shear has been measured out to a projected radius of 
3 $h_{65}^{-1}$ Mpc. We fit the average observed radial shear profile to a  
``generalized NFW profile'' on the form 

\begin{equation}
\rho(r) = \frac{\delta_c \rho_c}{(r/r_s)^\alpha (1+(r/r_s))^{3-\alpha}} .   
\label{eq:densprof} 
\end{equation}

For the model above, the characteristic density $\delta_c$ is 

\begin{equation}
\delta_c = \frac{200}{3} \left[\int_0^1 x^2 (c x)^{-\alpha}(1+c x)^{\alpha-3} dx \right]^{-1}.
\label{eq:deltc} 
\end{equation}

\begin{figure}
\begin{center}
\psfig{figure=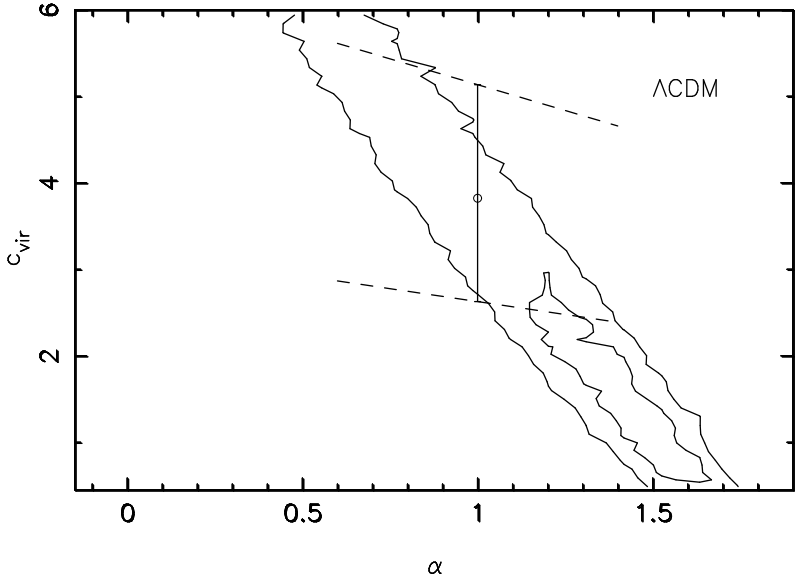,height=3in}
\end{center}
\caption{The contours show the $68\%$ and $95\%$ confidence intervals for the concentration $c_{\rm vir}$ and inner slope $\alpha$ of our average cluster halo. Also shown is the mean value and scatter in $c_{\rm vir}$ for an 
NFW halo of similar mass, predicted by Bullock et al.\ (2001). The dashed lines indicate lines along which the two parameters are degenerate. See also Dahle et al.\ (2003). 
\label{fig:Calpha}}
\end{figure}

\noindent
This model has a concentration parameter $c$ defined by $c = r_{200}/r_s$, 
where $\overline{\rho}(r_{200}) = 200 \rho_{\rm c}$ and $\rho_{\rm c}$ 
is the critical density. The outer slope at $r >> r_s$, $\rho \propto r^{-3}$, is  
chosen to be the same as the outer slope of simulated CDM halos, while the inner slope $\alpha$ 
is a free parameter. 
The result of our fit is given in Figure~\ref{fig:Calpha} which shows joint confidence 
limits for $\alpha$ and $c_{\rm vir}$ (defined as $c_{\rm vir} = r_{\rm vir}/r_s$, where 
$r_{\rm vir}$ is the virial radius of the halo). The CDM simulations predict a $(z+1)^{-1}$ 
redshift-dependence 
and significant intrinsic scatter in the values of $c_{\rm vir}$ (Jing 2000; Bullock et al.\ 2001). 
A prediction for the value and scatter of $c_{\rm vir}$ for massive ($M_{\rm vir} = 2 \times 10^{15} M_{\odot}$) clusters in a $\Lambda$CDM universe is indicated in Figure~\ref{fig:Calpha}.  
 
Our data constrain $\alpha$ to be in the range $0.9 < \alpha < 1.6$ (68\% CL), 
and $\alpha < 0.5$ is excluded at the 95\% level. We also find that the data are 
consistent with an isothermal sphere with a finite core, $\rho(r) \propto (r^2 + r_c^2)^{-1}$,   
where $r_c$ is the core radius. For the case of self-interacting dark matter, our constraints 
on the core radius implies a self-interaction cross section 
$\sigma_{\star} \le 0.1 {\rm cm}^2 {\rm g}^{-1}$ (c.f. Yoshida et al.\ 2000, Meneghetti et al.\ 2001). This is at least an order of magnitude 
smaller than the cross section required to explain the observed rotation curves of low surface 
brightness galaxies (Dav{\'e} et al.\ 2001). 

\section{Comparison with other results}\label{subsec:comp}

Different recent studies give a wide range of values for $\alpha$. 
Sand et al.\ (2003; see also Sand, these proceedings) use a combination of strong lensing 
data and spectroscopic measurements of stellar dynamics in the central galaxy of three  
clusters which contain both radial and tangential arcs to find an average value 
$\alpha = 0.52 \pm 0.05$, but they also find evidence for a significant scatter 
$\Delta \alpha \sim 0.3$. 
On the other hand, Bautz \& Arabadjis (2003) find $1 < \alpha < 2$ and 
Lewis, Buote \& Stocke (2003) find $\alpha = 1.19\pm 0.04$, based on {\em Chandra} 
observations of the X-ray luminous intracluster medium in four clusters and in one 
cluster, 
respectively. In contrast to our weak lensing study (which only probe the DM 
density profile at radii where the baryonic component is not dynamically dominant), these 
strong lensing and X-ray studies are not directly comparable to simulations that 
only contain collisionless CDM. The above results indicate that future observational studies should 
simultaneously take into account both the baryonic component in stars and in the X-ray 
luminous intracluster medium as well as the DM. Similarly, all these components must be 
properly modeled in numerical simulations, if the simulations are to be directly 
compared to cluster observations on small ($\le 10$\, kpc) scales. 
In any case, all the recent studies indicate that the core sizes of massive clusters are 
too small to be consistent with any self-interacting dark matter having a cross section 
large enough to explain the rotation curves of dwarf galaxies.
 
Like previous weak lensing studies (e.g., Clowe \& Schneider 2001, Hoekstra et al.\ 2002), we are 
not able to strongly distinguish between the outer slope of an isothermal 
sphere, $\rho \propto r^{-2}$, and the NFW slope $\rho \propto r^{-3}$.
However, in a recent work, Kneib et al.\ (2003) use a combination of weak and strong gravitational 
lensing data based on 
HST imaging of the cluster CL0024+17 to find an outer slope $>2.4$. Their data is adequately fit by a 
NFW profile with $c = 22^{+9}_{-5}$, significantly higher than typical observed concentration 
parameters of rich clusters (e.g., Hoekstra et al.\ 2002; Katgert, Biviano \& Mazure 2003), which 
are generally consistent with CDM predictions (see also Fig.~\ref{fig:Calpha}). 
However, {\em Chandra} X-ray data (Ota et al.\ 2003), as well as 
dynamical studies based on galaxy spectroscopy (Czoske et al.\ 2002), indicate that this is 
not a fully relaxed, spherically symmetric system. Weak lensing measurements of a representative 
sample of dynamically relaxed clusters out to even larger radii than we probe in our study should 
eventually settle the issue of the value of the outer slope. 

\section*{Acknowledgments}
I thank my collaborators Steen Hannestad and 
Jesper Sommer-Larsen, and acknowledge support from The Reseach Council of Norway
through a post-doctoral research fellowship.


\section*{References}
\begin{thebibliography}{99}

\bibitem{2003PhRvD..68b3511A} Arbey, 
A., Lesgourgues, J., \& Salati, P.\ 2003, \PRD, 68, 23511 

\bibitem{2003astro.ph..3313B} Bautz, M.~W.~\& Arabadjis, J.~S.\ 2003, ArXiv Astrophysics e-prints, 3313 

\bibitem{2001MNRAS.321..559B} Bullock, J.~S., Kolatt, T.~S., Sigad, Y., Somerville, R.~S., Kravtsov, A.~V., Klypin, A.~A., Primack, J.~R., \& Dekel, A.\ 2001, MNRAS, 321, 559

\bibitem{2003MNRAS.340..657D} de Blok, 
W.~J.~G., Bosma, A., \& McGaugh, S.\ 2003, MNRAS, 340, 657 
 
\bibitem{2001ApJ...546L..77C} Cen, R.\ 2001, ApJL, 546, L77 

\bibitem{2001A&A...379..384C} Clowe, D.~\& Schneider, P.\ 2001, A\& A, 379, 384

\bibitem{2002A&A...386...31C} Czoske, O., Moore, B., Kneib, J.-P., \& Soucail, G.\ 2002, A\& A, 386, 31.

\bibitem{2002ApJS..139..313D} Dahle, H., Kaiser, N., 
Irgens, R.~J., Lilje, P.~B., \& Maddox, S.~J.\ 2002, ApJS, 139, 313 

\bibitem{2003ApJ...588L..73D} 
Dahle, H., Hannestad, S., \& Sommer-Larsen, J.\ 2003, ApJL, 588, L73 

\bibitem{2001ApJ...547..574D} Dav{\' e}, R., Spergel, D.~N., Steinhardt, P.~J., \& Wandelt, B.~D.\ 2001, ApJ, 547, 574

\bibitem{2000NewA....5..103G} Goodman, J.\ 2000, New Astronomy, 5, 103

\bibitem{hs} Hannestad, S., \& Scherrer, R.J.\ 2000, \PRD, 62, 043522

\bibitem{2002MNRAS.333..911H} Hoekstra, H., Franx, M., Kuijken, K., \& 
van Dokkum, P.~G.\ 2002, MNRAS, 333, 911 

\bibitem{hbg} Hu, W., Barkana, R., \& Gruzinov, A.\ 2000, \PRL, 85, 1158

\bibitem{2000ApJ...535...30J} Jing, Y.~P.\ 2000, ApJ, 535, 30 

\bibitem{kkt} Kaplinghat, M., Knox, L., \& Turner, M.~S.\ 2000, \PRL, 85, 3335

\bibitem{2003astro.ph.10060K} Katgert, 
P., Biviano, A., \& Mazure, A.\ 2003, ArXiv Astrophysics e-prints, 10060 

\bibitem{1999ApJ...522...82K} Klypin, A., Kravtsov, A.~V., Valenzuela, O., \& Prada, F.\ 1999, ApJ, 522, 82

\bibitem{2003astro.ph..7299K} Kneib, J.~et al.\ 2003, 
ArXiv Astrophysics e-prints, 7299 

\bibitem{2003astro.ph..2036K} Kochanek, C.~S.~\& Dalal, N.\ 2003, ArXiv Astrophysics e-prints, 2036 

\bibitem{2003ApJ...586..135L} Lewis, A.~D., Buote, D.~A., \& Stocke, J.~T.\ 2003, ApJ, 586, 135 

\bibitem{2001MNRAS.325..435M} Meneghetti, M., Yoshida, N., Bartelmann, M., Moscardini, L., Springel, V., Tormen, G., \& White, S.~D.~M.\ 2001, MNRAS, 325, 435

\bibitem{2002PhRvD..66f3002M} 
Mohapatra, R.~N., Nussinov, S., \& Teplitz, V.~L.\ 2002, \PRD, 66, 63002 

\bibitem{1999ApJ...524L..19M} Moore, B., Ghigna, S., Governato, F., Lake, G., Quinn, T., Stadel, J., \& Tozzi, P.\ 1999a, ApJL, 524, L19

\bibitem{1999MNRAS.310.1147M} Moore, B., Quinn, T., Governato, F., Stadel, J., \& Lake, G.\ 1999b, MNRAS, 310, 1147

\bibitem{1997ApJ...490..493N} Navarro, 
J.~F., Frenk, C.~S., \& White, S.~D.~M.\ 1997, ApJ, 490, 493 

\bibitem{2003astro.ph..6580O} Ota, N., Pointecouteau, E., Hattori, 
M., \& Mitsuda, K.\ 2003, ArXiv Astrophysics e-prints, 6580 

\bibitem{2000ApJ...534L.127P} Peebles, P.~J.~E.\ 2000, 
ApJL, 534, L127 

\bibitem{pm} Piazza, F.~\& Marioni C.\ 2003, \PRL, 91, 141301 

\bibitem{2002ApJ...580L.129R} Robishaw, T., Simon, J.~D., \& Blitz, L.\ 2002, ApJL, 580, L129 

\bibitem{2003astro.ph..9465S} Sand, D.~J., Treu, T., Smith, G.~P., \& Ellis, R.~S.\ 2003, ArXiv Astrophysics e-prints, 9465 

\bibitem{2001ApJ...551..608S} Sommer-Larsen, J.~\& Dolgov, A.\ 2001, ApJ, 551, 608

\bibitem{ss} Spergel, D.~N.~\& Steinhardt P.~J.\ 2000, \PRL, 84, 3760

\bibitem{2000ApJ...544L..87Y} Yoshida, N., Springel, V., White, S.~D.~M., \&
 Tormen, G.\ 2000, ApJL, 544, L87 

\end{thebibliography}
\end{document}